\documentclass[11pt]{article}
\usepackage[letterpaper,margin=1in]{geometry}
\usepackage[T1]{fontenc}
\usepackage{lmodern}
\usepackage{amsmath,amssymb,amsthm,mathtools,microtype,enumitem}
\usepackage{array,booktabs,algorithm}
\usepackage[noend]{algpseudocode}
\algrenewcommand{\algorithmicrequire}{\textbf{Input}}
\usepackage{needspace}
\usepackage{flafter}
\usepackage[colorlinks=true,linkcolor=blue!45!black,citecolor=blue!45!black,urlcolor=blue!45!black]{hyperref}
\usepackage{xcolor}
\newtheorem{theorem}{Theorem}[section]
\newtheorem{lemma}[theorem]{Lemma}

\newtheorem{corollary}[theorem]{Corollary}
\theoremstyle{remark}
\newtheorem{remark}[theorem]{Remark}
\newcommand{\WED}{\operatorname{WED}}
\newcommand{\gap}{\bot}
\newcommand{\eps}{\varepsilon}
\newcommand{\softO}{\widetilde O}
\newcommand{\lce}{\operatorname{LCE}}
\newcommand{\Ext}{\operatorname{Ext}}
\newcommand{\Offer}{\textproc{Offer}}
\newcommand{\FAIL}{\mathsf{FAIL}}
\setlist{topsep=4pt,itemsep=3pt}
\hypersetup{pdftitle={Bounded Weighted Edit Distance with Metric Costs},pdfauthor={Jihan Wang}}
\title{Bounded Weighted Edit Distance with Metric Costs}
\author{Jihan Wang\\[0.3em]
    \small Siebel School of Computing and Data Science\\
    \small University of Illinois Urbana-Champaign\\
    \small Urbana, IL, USA\\
    \small \href{mailto:jihanw2@illinois.edu}{\nolinkurl{jihanw2@illinois.edu}}}
\date{}

\begin{document}
\maketitle

\begin{abstract}
\normalsize
We give a deterministic $O(n+k^2)$-time algorithm for bounded weighted
edit distance when the input costs form an integer metric on the
alphabet extended by a gap symbol. Here $n$ is the total input length
and $k$ is the given threshold on the distance. If the distance is at
most $k$, the algorithm returns it with an optimal alignment. Otherwise,
it reports that the distance exceeds $k$. The running time matches the
classical unit-cost bound and is independent of the alphabet size and of
the largest cost. The algorithm keeps only the farthest-reaching prefix
alignment for each cost and each difference between the total gap costs
of the aligned prefixes. For real metrics whose nonzero costs are at
least one, scaling and upward rounding give a
$(1+\varepsilon)$-approximation for the bounded problem in
$\widetilde{O}(n+(k/\varepsilon)^2)$ time. If insertions and deletions
have integer costs, the running time improves to
$\widetilde{O}(n+k^2/\varepsilon)$.
\end{abstract}

\section{Introduction}\label{sec:introduction}

Edit distance is the minimum number of insertions, deletions, and
substitutions needed to transform one string into another. The classical
dynamic program takes quadratic time~\cite{WF74}. In the bounded
problem, a threshold $k$ is also given. The task is to compute the
distance if it is at most $k$ and otherwise to report that it exceeds $k$.
The algorithm of Landau and Vishkin~\cite{LV88} combines the
diagonal-transition method~\cite{Ukk85,Mye86} with constant-time longest
common extension queries and solves the bounded problem in $O(n+k^2)$
time, where $n$ is the total input length.
The quadratic exponent of this bound is conditionally optimal for $k=n$,
since under the Orthogonal Vectors Hypothesis~\cite{Wil05} unit-cost
edit distance has no algorithm running in strongly subquadratic
time~\cite{BI18,BK15}.

Weighted edit distance assigns each edit a cost that depends on the
symbols involved. The dynamic program extends directly, but the
diagonal-transition method does not. It maintains a frontier of prefix
alignments and processes them in order of cost, which we call the
\emph{score}. A prefix alignment of the first $i$ symbols of one string
with the first $j$ symbols of the other ends at the vertex $(i,j)$ of
the edit grid. At each score, a unit-cost frontier keeps the farthest
vertex on each diagonal $i-j$. Under symbol-dependent gap costs, a
farther vertex on the same diagonal can leave a larger distance between
the remaining suffixes~\cite{DGHKS23,GK25}. This can happen even when
the costs form a metric. Keeping only the farther vertex can then
discard every optimal completion.

We replace the diagonal by a weighted analogue. The \emph{gap mass} of a
string is the total gap cost of its symbols, and the \emph{mass offset}
of a vertex $(i,j)$ is the gap mass of the first $i$ symbols of one
string minus that of the first $j$ symbols of the other. With unit costs
every gap costs one, so the mass offset is the diagonal $i-j$. By the
reverse triangle inequality, each edit changes the mass offset by at
most its cost~\cite{MTW26,DKK26}.

Our main lemma shows that removing prefixes of equal gap mass from both
strings cannot increase their distance. The substrings between two
vertices with the same mass offset have equal gap mass, so the farther
vertex leaves no larger distance between the remaining suffixes. For
unit costs, this recovers the monotonicity along diagonals used by the
diagonal-transition method. Hence the frontier needs to keep only the
farthest vertex at each score and mass offset. Since the absolute value
of the mass offset is at most the score, integer costs give $2c+1$
possible mass offsets at score $c$ and $O(k^2)$ frontier entries in
total. This gives the $O(n+k^2)$ bound for integer metric costs.

\Needspace{12\baselineskip}
\subsection{Our results}\label{sec:results}

Let $X$ and $Y$ be strings of total length $n$ over an alphabet
$\Sigma$, and let $\gap\notin\Sigma$ be a gap symbol. The input metric
$w$ is defined on $\Sigma\cup\{\gap\}$. Inserting or deleting $a$ costs
its \emph{gap cost} $w(a,\gap)$, and substituting $b$ for $a$ costs
$w(a,b)$. The weighted edit distance $\WED_w(X,Y)$ is the minimum total
cost of a sequence of edits that transforms $X$ into $Y$. We call a cost
function $w$ on $\Sigma\cup\{\gap\}$ \emph{normalized} if $w(a,a)=0$ for
every symbol $a$ and $w(a,b)\ge1$ for all distinct symbols $a$ and $b$.

We work in the real RAM model, which supports exact arithmetic and
comparisons on cost values but no rounding to integers. We assume that
symbols are integers bounded by a polynomial in $n$ and that $w$ is
given by a constant-time oracle. For an integer metric, the oracle
returns integers that support constant-time arithmetic and array
indexing.

\begin{theorem}\label{thm:integer}
Let $w$ be an integer metric and let $k\ge1$ be an integer threshold.
There is a deterministic algorithm that returns $\WED_w(X,Y)$ and an
optimal alignment if $\WED_w(X,Y)\le k$ and otherwise reports that the
distance exceeds $k$. The algorithm uses $O(n+k^2)$ time and space.
\end{theorem}

The bound is independent of the alphabet size and of the largest cost.
The $O(k^2)$ frontier size relies on integer scores and mass offsets.
For normalized real metrics, we scale the costs and round them up to
integers to restore this structure.

\begin{theorem}\label{thm:approximation}
Let $w$ be a normalized real metric and let $k\ge1$ be an integer
threshold. For every $0<\eps\le1$, there is a deterministic algorithm
that either reports that the distance exceeds $k$ or returns a value $A$
and an alignment of cost at most $A$ such that
\[
 \WED_w(X,Y)\le A\le(1+\eps)\WED_w(X,Y).
\]
If $\WED_w(X,Y)\le k$, it returns such a value and alignment. The
algorithm uses $O(n+(k/\eps)^2)$ space and $O(n+(k/\eps)^2\log(k/\eps))$
time.
\end{theorem}

The notation $\softO$ suppresses factors polylogarithmic in $n$, $k$,
and $1/\eps$. For normalized real metrics with integer gap costs, the
running time improves to $O(n+(k^2/\eps)\log(k/\eps))$.

\begin{remark}
In concurrent and independent work, Gorbachev~\cite{Gor26} obtains a
deterministic $(1+\eps)$-approximation in
$\softO(n+k^2/\sqrt{\eps})$ time for bounded weighted edit distance with
general normalized costs. This result has a better dependence
on $\eps$ and does not require the metric assumption.
\end{remark}

Table~\ref{tab:comparison} compares our bounds with those of Gorbachev
and Kociumaka~\cite{GK25}, which allow general costs.

\begin{table}[H]
\centering
\begin{tabular}{@{}>{\raggedright\arraybackslash}p{0.43\linewidth}ll@{}}
\toprule
Cost model
 & \begin{tabular}[b]{@{}l@{}}Previous bound~\cite{GK25}\\(general costs)\end{tabular}
 & \begin{tabular}[b]{@{}l@{}}This paper\\(metric costs)\end{tabular}\\
\midrule
Positive integer costs at most $W$
 & $\softO(n+Wk^2)$ & $O(n+k^2)$\\
Arbitrary positive integer costs
 & $\softO(n+k^{5/2})$ & $O(n+k^2)$\\
Normalized real costs, $(1+\eps)$-approximation
 & $\softO(n+(k/\eps)^{5/2})$
 & $\softO(n+(k/\eps)^2)$\\
\bottomrule
\end{tabular}
\caption{Running times for bounded weighted edit distance with threshold $k\ge1$.}
\label{tab:comparison}
\end{table}

\Needspace{7\baselineskip}
\subsection{Related work}\label{sec:related}

Das, Gilbert, Hajiaghayi, Kociumaka, and Saha~\cite{DGHKS23} give an
$O(n+k^5)$-time algorithm for general normalized costs. Cassis,
Kociumaka, and Wellnitz~\cite{CKW23} improve this to
$\softO(n+\sqrt{n}\,k^{3/2})$ time and show that this bound is conditionally
optimal for $\sqrt n\le k\le n$. Their lower-bound constructions include
metric rational costs with input-dependent denominators.

Gorbachev and Kociumaka~\cite{GK25} obtain the previous bounds in
Table~\ref{tab:comparison}. For integer costs at most $W$, their
algorithm uses a hierarchical alignment data structure and core-sparse
Monge matrices. For a fixed finite alphabet and a fixed rational cost
function, they also obtain quadratic dependence on $k$. In contrast, our
bound allows the integer metric to be part of the input. The metric
assumption lets our algorithm use only longest common extension queries
and a frontier of prefix alignments.

Boneh, Gorbachev, and Kociumaka~\cite{BGK25} give dynamic algorithms
and conditional lower bounds for bounded weighted edit distance under
general normalized costs. Charalampopoulos, Kociumaka, and
Wellnitz~\cite{CKW25} study pattern matching under weighted edit
distance, including metrics with bounded integer costs.

For weighted edit distance with arbitrary metric costs,
Kuszmaul~\cite{Kus19} gives a randomized $O(n^{\gamma})$-approximation in
$\softO(n^{2-\gamma})$ time for every $0<\gamma<1$ on strings of equal
length. Mader, Tavasoli, and Wang~\cite{MTW26} and Das, Kipouridis, and
Kociumaka~\cite{DKK26} give randomized $(3+\eps)$-approximations in
strongly subquadratic time. Both papers use prefix-mass potentials such
as our mass offset to restrict candidate locations. We use the same
potential to show that the farthest vertex at each score and mass offset
dominates all others.

\section{Preliminaries}\label{sec:preliminaries}

\paragraph{Strings and alignments.}
Let $m=|X|$ and $r=|Y|$, and write $g(a)=w(a,\gap)$ for the gap cost of
$a$. We index positions from zero and write $Z[i,j)$ for the substring
$Z[i]\cdots Z[j-1]$ and $Z[i{:}]=Z[i,|Z|)$ for a suffix.

For a fixed metric $w$, write $D(U,V)=\WED_w(U,V)$ and $D=D(X,Y)$.
The \emph{edit grid} (or alignment graph) of $U$ and $V$ has vertex set
$\{0,\ldots,|U|\}\times\{0,\ldots,|V|\}$ and the following edges.
\[
\begin{array}{lll}
 (i,j)\longrightarrow(i+1,j)
   &\text{at cost }g(U[i])&\text{(deletion)},\\
 (i,j)\longrightarrow(i,j+1)
   &\text{at cost }g(V[j])&\text{(insertion)},\\
 (i,j)\longrightarrow(i+1,j+1)
   &\text{at cost }w(U[i],V[j])&\text{(substitution or match)}.
\end{array}
\]
An \emph{alignment} is a path from $(0,0)$ to $(|U|,|V|)$, and a
\emph{prefix alignment} is a path from $(0,0)$ to any vertex. A diagonal
edge with $U[i]=V[j]$ is a \emph{match}, and every other edge is an
\emph{edit}. Since $w$ is a metric, matches cost zero and edits have
positive cost. By the triangle inequality, weighted edit distance equals
the minimum alignment cost~\cite{WF74,Sel74}. Every subpath of an
optimal alignment is a minimum-cost path between its endpoints.

\paragraph{Gap masses.}
For a string $Z$ and $0\le i\le|Z|$, define
\[
 \mu(Z)=\sum_{s=0}^{|Z|-1}g(Z[s]),\qquad
 \mu_Z(i)=\mu(Z[0,i)).
\]
For a vertex $(i,j)$ of the edit grid of $X$ and $Y$, let
$\delta(i,j)=\mu_X(i)-\mu_Y(j)$. We call $\mu(Z)$ the \emph{gap mass} of
$Z$, $\mu_Z(i)$ a \emph{prefix mass}, and $\delta(i,j)$ the
\emph{mass offset}. The functions $\mu_X$ and $\mu_Y$ are strictly
increasing because $g(a)>0$ for every $a\in\Sigma$. The reverse triangle
inequality gives
\begin{equation}\label{eq:gap-inequalities}
 |g(a)-g(b)|\le w(a,b).
\end{equation}
We recall the following bound on mass offsets from~\cite{MTW26,DKK26}.

\begin{lemma}\label{lem:potential}
A path from $(i,j)$ to $(i',j')$ of cost $c$ in the edit grid of $X$
and $Y$ satisfies
\[
 |\delta(i',j')-\delta(i,j)|\le c.
\]
In particular, every prefix alignment of cost $c$ ends at a vertex
$(i,j)$ with $|\delta(i,j)|\le c$.
\end{lemma}
\begin{proof}
Deleting $a$ increases the mass offset by $g(a)$, and inserting $b$
decreases it by $g(b)$. A diagonal edge pairing $a$ with $b$ changes it by
$g(a)-g(b)$. The absolute value of this change is at most $w(a,b)$
by~\eqref{eq:gap-inequalities}. Summing the absolute changes along
the path proves the claim.
\end{proof}

\paragraph{Longest common extensions.}\label{sec:lce}
Let $\lce(i,j)$ be the length of the longest common prefix of
$X[i{:}]$ and $Y[j{:}]$, and let
\[
 \Ext(i,j)=(i+\lce(i,j),j+\lce(i,j))
\]
be the vertex reached from $(i,j)$ by matching this prefix. Let
$\#\notin\Sigma$ be a separator symbol. The suffix array~\cite{KSB06} of
$X\#Y$ sorts its suffixes lexicographically, and the longest-common-prefix
array~\cite{KLAAP01} stores the common-prefix lengths of consecutive
suffixes in this order. The common-prefix length of two different
suffixes is the minimum of this array between their ranks, which a
range-minimum structure~\cite{BF00} returns in constant time. Because $Y$
contains no $\#$, the suffixes that start at $i$ and at $m+1+j$ have
common-prefix length $\lce(i,j)$ for $j<r$. Since symbols are polynomially
bounded integers, building the resulting \emph{LCE index} takes $O(n)$
time and space.

\section{Prefix-mass dominance}\label{sec:dominance}

We show that removing prefixes of equal gap mass from two strings
cannot increase their distance. The same argument recovers the known
common-prefix cancellation property~\cite{CKW23}.

\begin{lemma}\label{lem:removal}
For all strings $P,P',U,V$,
\[
 D(U,V)\le D(PU,P'V)+|\mu(P)-\mu(P')|.
\]
\end{lemma}
\begin{proof}
Let $(i,j)$ be the first vertex on an optimal alignment of $PU$ and
$P'V$ with $i\ge|P|$ and $j\ge|P'|$. Since each coordinate increases by
at most one per edge, $i=|P|$ or $j=|P'|$.

Suppose $i=|P|$ and let $c$ be the cost of the subpath up to $(i,j)$,
which aligns $P$ with $P'S$ for a prefix $S$ of $V$.
Applying Lemma~\ref{lem:potential} to $PU$ and $P'V$ gives
\[
 |\mu(P)-\mu(P')-\mu(S)|\le c
 \quad\text{and hence}\quad
 \mu(S)\le c+|\mu(P)-\mu(P')|.
\]

The remaining subpath aligns $U$ with $V[|S|{:}]$ at cost
$D(PU,P'V)-c$. Inserting $S$ at cost $\mu(S)$ and then following this
subpath gives an alignment of $U$ with $V$ of cost at most
\[
 D(PU,P'V)-c+\mu(S)
 \le D(PU,P'V)+|\mu(P)-\mu(P')|.
\]
When $j=|P'|$, the symmetric construction deletes a prefix of $U$.
\end{proof}

\begin{corollary}\label{cor:dominance}
Let $(i_1,j_1)$ and $(i_2,j_2)$ be vertices of the edit grid of $X$ and
$Y$. If $i_2\ge i_1$ and $\delta(i_2,j_2)=\delta(i_1,j_1)$, then
$j_2\ge j_1$ and
\[
 D(X[i_2{:}],Y[j_2{:}])\le D(X[i_1{:}],Y[j_1{:}]).
\]
\end{corollary}
\begin{proof}
Equality of the mass offsets gives
$\mu_Y(j_2)-\mu_Y(j_1)=\mu_X(i_2)-\mu_X(i_1)\ge0$.
Since $\mu_Y$ is strictly increasing, $j_2\ge j_1$.
The substrings $X[i_1,i_2)$ and $Y[j_1,j_2)$ have equal gap mass.
Lemma~\ref{lem:removal} with $P=X[i_1,i_2)$, $P'=Y[j_1,j_2)$,
$U=X[i_2{:}]$, and $V=Y[j_2{:}]$ gives the distance inequality.
\end{proof}

\begin{corollary}\label{cor:cancellation}\label{lem:extension}
For all strings $P,U,V$, $D(PU,PV)=D(U,V)$. Applying $\Ext$ also
preserves both the mass offset and the distance between the remaining
suffixes.
\end{corollary}
\begin{proof}
Lemma~\ref{lem:removal} with $P'=P$ gives $D(U,V)\le D(PU,PV)$, and
matching the two copies of $P$ for free gives the reverse inequality.
The map $\Ext$ skips a common prefix of the remaining suffixes, so both
prefix masses increase by its gap mass and the mass offset is unchanged.
By the first claim, the distance between the remaining suffixes is
unchanged as well.
\end{proof}

\section{The bounded weighted frontier}\label{sec:integer}\label{sec:frontier}

Let $w$ be an integer metric. The frontier processes prefix alignments
by increasing score. A \emph{candidate} is a vertex together with the
score of a prefix alignment that reaches it. Before storing a candidate,
the frontier replaces its vertex $(i,j)$ by $\Ext(i,j)$. We call this step
\emph{match extension}, and it ensures that every stored vertex $(i,j)$
satisfies $\lce(i,j)=0$.

Lemma~\ref{lem:potential} places the mass offset of every candidate of
score $c$ in $[-c,c]$. For each integer score $0\le c\le k$ and mass
offset $-c\le d\le c$, the entry $R[c,d]$ stores the farthest candidate
of score $c$ and mass offset $d$ found so far.
Corollary~\ref{cor:dominance} shows that at the same mass offset a
larger $i$ never comes with a smaller $j$, so the farthest candidate is
the one with the largest $i$. The entries $R[c,\cdot]$ form
\emph{score layer}~$c$. For traceback, $R[c,d]$ also stores the
\emph{predecessor entry} $p$ from which its candidate was generated.

An edge from a stored vertex $(i,j)$ to $(i',j')$ changes the mass
offset by $\mu(X[i,i'))-\mu(Y[j,j'))$, and match extension leaves it
unchanged by Corollary~\ref{cor:cancellation}. The frontier therefore
computes the mass offset of each new candidate from that of the stored
candidate.

Algorithm~\ref{alg:frontier} gives the frontier and its procedure
$\Offer$. The first call of $\Offer$ stores the \emph{root} candidate,
which has score $0$ and vertex $\Ext(0,0)$. Following predecessor
entries from a stored candidate back to the root candidate gives its
\emph{predecessor chain}.

\begin{algorithm}[t]
\caption{Bounded weighted edit distance}\label{alg:frontier}
\begin{algorithmic}[1]
\Require strings $X$, $Y$, integer metric $w$, threshold $k\ge1$, and
the LCE index
\Procedure{Offer}{$c,i,j,d,p$}
  \If{$c\le k$}
    \State $(i,j)\gets\Ext(i,j)$
    \If{$R[c,d]$ is empty or $R[c,d].i<i$}
      \State $R[c,d]\gets(i,j,p)$
    \EndIf
  \EndIf
\EndProcedure
\Function{Frontier}{$k$}
  \State Initialize $R[c,d]$ as empty for $0\le c\le k$, $-c\le d\le c$
  \State \Call{Offer}{$0,0,0,0,\mathsf{null}$}
  \For{$c=0,1,\ldots,k$}
    \For{$d=-c,-c+1,\ldots,c$ with $R[c,d]$ nonempty}
      \State $(i,j)\gets(R[c,d].i,R[c,d].j)$
      \If{$(i,j)=(m,r)$}
        \State \Return $c$ and the alignment from the predecessor chain
      \EndIf
      \ForAll{outgoing edges $(i,j)\to(i',j')$ of cost $t$}
        \State $d'\gets d+\mu(X[i,i'))-\mu(Y[j,j'))$
        \State \Call{Offer}{$c+t,i',j',d',(c,d)$}
      \EndFor
    \EndFor
  \EndFor
  \State \Return $\FAIL$
\EndFunction
\end{algorithmic}
\end{algorithm}

\Needspace{8\baselineskip}
\subsection{Correctness and running time}\label{sec:traceback}

\begin{proof}[Proof of Theorem~\ref{thm:integer}]
We analyze Algorithm~\ref{alg:frontier}. Match extension makes every
outgoing edge of a stored vertex an edit of positive cost, so no score
layer changes once its processing begins.
Thus the candidate stored in a predecessor entry never changes after it
generates a new candidate. The prefix alignment of the root
candidate matches the longest common prefix of $X$ and $Y$, and each
later stored candidate extends the prefix alignment of the candidate in
its predecessor entry by one edit followed by matches. Hence the
predecessor chain of every stored candidate represents a prefix
alignment whose cost equals the score of the candidate.

\emph{Correctness.} Suppose $D\le k$, and call a candidate of score $c$
at vertex $(i,j)$ \emph{feasible} if
\[
 c+D(X[i{:}],Y[j{:}])\le D.
\]
A feasible candidate can be completed to an optimal alignment, and its
score is at most $D$. A stored candidate is replaced only by one at a
farther vertex with the same score and mass offset, so feasibility is
preserved by Corollary~\ref{cor:dominance}. Hence a score layer that
holds a feasible candidate keeps one until it is processed.

Every stored candidate at $(m,r)$ represents an alignment and hence has
score at least $D$, so the algorithm does not return before score layer
$D$. Suppose that score layer $c$ holds a feasible candidate at a vertex
$(i,j)\ne(m,r)$ when its processing begins. Match extension gives
$\lce(i,j)=0$, so an optimal alignment of $X[i{:}]$ and $Y[j{:}]$ starts
with an edit. If this edit has cost $t$ and ends at $(i',j')$, then
\[
 t+D(X[i'{:}],Y[j'{:}])=D(X[i{:}],Y[j{:}]).
\]
Combined with feasibility this gives $c+t+D(X[i'{:}],Y[j'{:}])\le D$. In
particular $c<c+t\le D\le k$, so the algorithm processes this candidate
and offers $(i',j')$ at score $c+t$. By Corollary~\ref{cor:cancellation},
the candidate at score $c+t$ and vertex $\Ext(i',j')$ is feasible. By
Corollary~\ref{cor:dominance}, so is the candidate stored in its entry.
Hence score layer $c+t$ also holds a feasible candidate when its
processing begins.

Score layer $0$ holds the root candidate, which is feasible by
Corollary~\ref{cor:cancellation}. Repeating this argument increases the
score but keeps it at most $D$, so we reach a score layer that holds a
feasible candidate at $(m,r)$. This candidate represents an alignment of
cost at most $D$, so its score is $D$. The algorithm therefore returns
$D$ and an optimal alignment while processing score layer $D$. If $D>k$,
the algorithm returns $\FAIL$ because no stored candidate at $(m,r)$ has
score at most $k$.

\emph{Traceback.} The algorithm recovers the alignment from the
predecessor chain of the stored candidate at $(m,r)$. Consider a
candidate at $(i,j)$ on this chain whose predecessor entry stores the
vertex $(u,v)$. Matches preserve the diagonal $i-j$, so its change
identifies the edit between the two vertices. It is a deletion of $X[u]$
if $(i-u)-(j-v)=1$, an insertion of $Y[v]$ if $(i-u)-(j-v)=-1$, and a
substitution of $Y[v]$ for $X[u]$ otherwise.

\emph{Running time.} The frontier has $O(k^2)$ entries because score $c$
allows $2c+1$ mass offsets. Each entry is processed once and generates
at most three candidates, and each candidate takes one LCE query and
constant additional work. Generating the candidates of one entry takes
$O(1)$ queries to $w$, so the algorithm makes $O(k^2)$ queries. The
predecessor chain has at most $k$ edits, and expanding the matches takes
$O(n)$ time. Building the LCE index takes $O(n)$ time and space. In
total, the algorithm uses $O(n+k^2)$ time and space.
\end{proof}

\section{Normalized real metrics}\label{sec:approximation}\label{sec:extensions}

Upward rounding lets us apply the algorithm of Theorem~\ref{thm:integer}
to normalized real metrics. Integer gap costs give fewer mass offsets
after rounding.

\subsection{Upward rounding}

Let $w$ be a normalized real metric and let $0<\eps\le1$. We scale the
costs by a power of two $Q$ with $1/\eps\le Q<2/\eps$. Because the model
has no rounding primitive, we find $Q$ by starting from $Q=1$ and
doubling until $Q\ge1/\eps$. We define the rounded costs
\[
 w_Q(a,b)=\lceil Qw(a,b)\rceil.
\]
The rounded costs form an integer metric. The sum
$\lceil Qw(a,b)\rceil+\lceil Qw(b,b')\rceil$ is an integer that is at
least $Qw(a,b')$ by the triangle inequality for $w$, so it is at least
$\lceil Qw(a,b')\rceil$.

Every alignment has rounded cost at least $Q$ times its original cost.
An optimal alignment for $w$ has at most $D$ edits, since each costs at
least one. Rounding each scaled edit cost upward adds less than one.
Therefore the rounded distance $D_Q=\WED_{w_Q}(X,Y)$ satisfies
\begin{equation}\label{eq:rounding-guarantee}
 QD\le D_Q\le(Q+1)D.
\end{equation}
The estimate $A=D_Q/Q$ satisfies $D\le A\le(1+\eps)D$. An optimal
alignment for $w_Q$ has original cost at most $A$.

Since the frontier explores only prefix alignments of cost at most its
threshold, we can cap rounded costs at an integer above the threshold.
Capping also bounds the search range when we evaluate rounded costs,
which keeps the running time independent of the largest cost. For an
integer cap $C>0$, define
\[
 w_{Q,C}(a,b)=\min\{w_Q(a,b),C\}.
\]
Capping preserves the triangle inequality because
$\min\{x+y,C\}\le\min\{x,C\}+\min\{y,C\}$ for $x,y\ge0$. An alignment of
cost below $C$ under either metric has the same cost under both because
it cannot use an edge capped to $C$.

\paragraph{Rounded queries.}\label{sec:rounded-queries}
Without a rounding primitive, we evaluate $w_{Q,C}(a,b)$ by exact
comparisons and return zero if $a=b$. Otherwise, we query $z=w(a,b)$
and find the largest integer $q$ in $[0,C]$ strictly below $Qz$. We
return $\min\{q+1,C\}$, which equals $\min\{\lceil Qz\rceil,C\}$.

Starting from $q=0$, we scan powers of two $h$ from the largest one at
most $C$ down to $1$ and add $h$ to $q$ when $q+h\le C$ and $q+h<Qz$.
This binary search over the bits of $q$ takes $O(\log C)$ exact
operations.

\begin{proof}[Proof of Theorem~\ref{thm:approximation}]
We run the algorithm of Theorem~\ref{thm:integer} on $w_{Q,C}$ with
threshold $B=(Q+1)k$ and cap $C=B+1$. Since $B<C$, a successful call
returns the rounded distance $D_Q$. We return $A$ with the recovered
alignment, whose guarantees follow from~\eqref{eq:rounding-guarantee}.
If $D\le k$, then~\eqref{eq:rounding-guarantee} gives $D_Q\le B$ and the
call succeeds. A failed call therefore implies $D>k$.

On $w_{Q,C}$, the algorithm of Theorem~\ref{thm:integer} performs
$O(n+B^2)$ operations and makes $O(B^2)$ rounded queries.
Each rounded query takes $O(\log C)$ exact operations. Choosing $Q$
takes $O(\log(1/\eps))$ doublings. Since $B$ and $C$ are $O(k/\eps)$,
the total time is $O(n+(k/\eps)^2\log(k/\eps))$ and the space is
$O(n+(k/\eps)^2)$.
\end{proof}

\subsection{Integer gap costs}

Suppose the original gap costs are integers. Then
$w_Q(a,\gap)=Qw(a,\gap)$, so every mass offset under $w_Q$ is a multiple
of $Q$. Rounded substitution costs need not be multiples of $Q$, so we
still index scores by all integers. Iterating only over mass offsets
that are multiples of $Q$ reduces the frontier size by a factor of $Q$.

\begin{corollary}\label{cor:integer-gaps}
For normalized real metrics with integer gap costs, the guarantees of
Theorem~\ref{thm:approximation} can be achieved in $O(n+k^2/\eps)$ space
and $O(n+(k^2/\eps)\log(k/\eps))$ time.
\end{corollary}
\begin{proof}
We use threshold $B=(Q+1)k$ and cap $C=Q(2k+1)>B$. The bounds
in~\eqref{eq:rounding-guarantee} give the same approximation and
threshold guarantees. Since $C$ is a multiple of $Q$, all capped gap
costs and mass offsets remain multiples of $Q$. We therefore store the
frontier entries as $R[c,d/Q]$.

There are $B+1=O(Qk)$ score layers. In score layer $c$,
Lemma~\ref{lem:potential} gives $|d|\le c\le B$. The spacing $Q$ leaves
at most $2B/Q+1=O(k)$ possible mass offsets, so the frontier has
$O(Qk^2)$ entries. The algorithm therefore performs $O(n+Qk^2)$
operations and makes $O(Qk^2)$ rounded queries. Each rounded query
takes $O(\log C)=O(\log(k/\eps))$ exact operations, and $Q=O(1/\eps)$
gives the stated bounds.
\end{proof}

\section*{Acknowledgments}
In~\cite{MTW26}, Ethan Mader, Borna \mbox{Tavasoli}, and the author use
prefix-mass potentials in an approximation algorithm. The present paper
arose from the author's observation that the same potential supports an
exact dominance rule for bounded weighted edit distance. The author
thanks Tomasz Kociumaka for discussions of this idea. OpenAI Codex and
Anthropic Claude Code assisted only with
manuscript revision, mathematical and bibliographic checks, and
typesetting. The author assumes responsibility for all content.

\begingroup
\raggedright
\bibliographystyle{alphaurl}
\bibliography{ref}
\endgroup

\end{document}